# CONCEPTUAL STEPS TOWARDS EXPLORING THE FUNDAMENTAL NATURE OF OUR SUN

**Attila Grandpierre**[*]

Konkoly Observatory
Budapest, Hungary

Motto: "*The purpose of life is the investigation of the Sun, the Moon, and the heavens.*" - Anaxagoras 459 BC

## SUMMARY

One of the basic questions of solar research is the nature of the Sun. We show here how the plasma nature of the Sun leads to the self-generation of solar activity. The release of magnetic, rotational, gravitational, nuclear energies and that of the gravity mode oscillations deviate from uniformity and spherical symmetry. Through instabilities they lead to the emergence of sporadic and localized regions like flux tubes, electric filaments, magnetic elements and high temperature regions. A systematic approach exploring the solar collective degrees of freedom, extending to ordering phenomena of the magnetic features related to Higgs fields, is presented.

Handling solar activity as transformations of energies from one form to another one presents a picture on the network of the energy levels of the Sun, showing that the Sun is neither a mere "ball of gas" nor a "quiescent steady-state fusion-reactor machine", but a complex self-organizing system. Since complex self-organizing systems are similar to living systems (and, by some opinion, identical with them), we also consider what arguments indicate the living nature of the Sun. Thermodynamic characteristics of the inequilibrium Sun are found important in this respect and numerical estimations of free energy rate densities and specific extropy flows are derived.

## CLASSIFICATION

*Corresponding author, $\eta$: grandp@konkoly.hu; +36 1 391 9360;
Konkoly Observatory of the Hungarian Academy of Sciences, H-1525 Budapest, P.O.Box 67;
www.konkoly.hu/staff/grandpierre

## INTRODUCTION

Solar physicists would like to know as much from the Sun as possible, as thoroughly as possible, as validly as possible. Following this aim, the solar physicists' community looked forward to find new windows for the Sun. The entrance of helioseismology or neutrino detectors resulted in a significant improvement in the study of the Sun. Regarding theoretical models, a common fundamental assumption contrasting with present day approaches is to regard the Sun as a mere gaseous body. The Sun is the nearest star, and a "star" is "a luminous ball of gas" [48]. At the same time, we know that the Sun is not in a gaseous, but in a plasma state. In the solar interior, the temperature is so high that matter is ionized and therefore the whole interior of the Sun is in a plasma state, including the solar core. Plasma consists of electrically charged particles that respond collectively to electromagnetic force. Because of their strong interaction with electromagnetism, plasmas display a complexity in structure and motion that far exceeds that found in matter in the gaseous, liquid, or solid states. The electromagnetic interaction is 39 orders of magnitude stronger than the gravitational one, and, correspondingly, it is enormously richer in nonlinear interactions. The collective plasma processes are associated in particular with various plasma instabilities. As a rule the development of instability is accompanied by an increase in the electric field strength, which can attain large values. Consequently, even in the absence of intense external fields, relatively strong fields can still occur spontaneously in a plasma due to the growth of instability [61]. Plasma microinstabilities are localized, usually high frequency phenomena that cannot be described in MHD but in the kinetic models. In general, the plasma can support electric currents (e.g. [26]). Plasmas are extremely complicated physical systems fundamentally different from classic neutral gases, especially when there is a magnetic field present, and even more so when gravitational and nuclear energy sources liberate energies in a strong dependence on temperature.

Moreover, the main body of the Sun, the radiative interior of this "luminous ball of gas," is regarded as being quiet, with no other changes occurring beside the slow transmutation of elements due to nuclear reactions. Bahcall [2] characterized a general view with the following sentence: "The Sun's interior is believed to be in a quiescent state and therefore the relevant physics is simple". On the other hand, the "plasma diagram" showing the ranges of temperature and number density of natural and man-made plasmas posits the solar core in line with terrestrial lightning, but at higher density and temperature (see http://fusedweb.pppl.gov/CPEP/chart.html).

The idea that the solar core may show dynamism is an old one; but recently it has received theoretical and observational support (see in [30, 32]). There is a whole range of interactions liberating energies in the solar radiative interior, but it was assumed that these energies are negligible. Within the most important energy sources, we mention here the nuclear, gravitational, rotational, magnetic, oscillatory energies, p-mode, g-mode, r-mode waves and tidal waves. In the actual solar radiative interior a wide range of magnetic, rotational, tidal and oscillatory instabilities are present and the solar core is changing on a wide variety of timescales. Magnetohydrodynamic calculations (e.g. [58, 60]) indicated the general presence of MHD instabilities in the solar radiative zone. Electric conductivity in the solar plasma is very high. Solar core is widely penetrated by dynamic influences that generate movements within it. Such motions are the electric currents, tidal waves, g-mode oscillations, as well as the ones triggered by different instabilities like e.g. the magnetic, rotational [56], and baroclinic [57] ones. Any slight hydrodynamic change of the plasma generates electromagnetic disturbances. Plasma instabilities have a local and nonlinear character. When the magnetic field is unstable, the magnetic energy is transformed into other energy forms like into the electric energy of currents. Since the initial time development of magnetic instabilities has an exponential character in general, the related electric currents grow exponentially until they reach intensity where the pinch-effect contracts the currents and filamentary structures will be formed. These current filaments in the solar core may stabilize themselves and lead to significant local heating. Filamentation is a general, well-known property of plasma [11, 43].

Despite of its plasma state, the solar core is widely regarded as being in a gaseous state. Plasma nature of the solar core has been ignored because of the general assumption that the solar core is quiescent and, therefore, being static, the solar core does not show any peculiar plasma behaviour. In the solar core a lot of different types of motions set up [30]. Therefore plasma processes will also set up and they will lead to significant dynamic phenomena. In the solar core the gaseous pressure ($\sim 10^{12}$ N/m$^2$) is much larger than the magnetic pressure since the average toroidal field has a helioseismic upper limit of $3 \cdot 10^3$ T [62]. Nevertheless, the criterion for neglecting magnetic effects in the treatment of a problem in gas dynamics is that the Lundquist parameter $L_u = \mu^{1/2} \sigma B l_c / \rho_m^{1/2}$ (measuring the ratio of the magnetic diffusion time to the Alfven travel time, where $m$ is the magnetic permeability $4\pi \cdot 10^{-7}$ H·m$^{-1}$, $\sigma$ is the electric conductivity in siemens/m, $B$ is the strength of the magnetic field in Tesla, $l_c$ is a characteristic length of the plasma in meter, and $\rho_m$ is the mass density in kg/m$^3$) is much less than unity, $L_u \ll 1$ [43, p. 19]. Now for the solar core $\sigma \sim 10^7$ S/m, $B \gg 10^{-7}$ to $3 \cdot 10^3$ T, $l_c \sim 10^8$ m, $\rho_m \sim 10^5$ kg/m$^3$, and so $L_u \sim 10^3$ to $3 \cdot 10^{13}$. Therefore, plasma effects may play a dominant role in the dynamism of the solar core. We note that even when $L_u \ll 1$, the hydrodynamic movements may amplify the magnetic fields to values to $L_u \gg 1$ later on.

The nature of solar rotational instabilities is far from being understood (see e.g. [56]). Solar rotation during the last $4,6 \cdot 10^9$ years has been spun down from $E_{rot, 0} \sim 10^{38}$ J to the present one, representing $E_{rot, present} \sim 2,4 \cdot 10^{35}$ J [1]. Charbonneau and MacGregor [12, 13] calculated the solar spin-down from the zero-age main sequence. From their Fig. 2a, one can read that the present rate of solar spin-down corresponds to $(\Delta E_{rot}/\Delta t)_{present} \sim 2 \cdot 10^{27}$ J/year. Although Couvidat et al. [16] obtained a flat rotation profile down to $0,2 \cdot R_{Sun}$, the uncertainties in the rotation rate are still quite large below $0,3 \cdot R_{Sun}$. It is widely thought that the deceleration of the solar core is due to magnetic breaking. The flat rotation curve at 430 nHz in the solar radiative interior lends support for a magnetic field strong enough to suppress any differential rotation might arise from angular momentum redistribution through the gravity waves [59]. It is well known that the dissipation of rotational energy is used to drive the dynamo, and, in general, solar activity in the solar envelope. At the same time, the main part of the solar rotational energy is dissipated in the core as it is decelerated. It may seem plausible that the rotational energy is dissipated in the solar core, too, in a form suitable to drive activity phenomena.

Gravitational energy liberation in the solar core deviates from perfect uniformity and homogeneity since the solar core does not show perfect spherical symmetry. Asphericity may lead to the development of sites where the gravitational energy liberation is enhanced relative to the environment, and so a local heated region may develop, which will expand and become lighter.

All aspects of solar activity (Ellerman bombs, bright points, flares, particle acceleration etc.) are self-similar and their statistical behaviour follows well-defined power laws. This last point reinforces the belief that the solar atmosphere is coupled, through the magnetic field, with the convection zone [66]. Regarding that the solar atmosphere is widely different from the interior of the Earth, and despite this difference similar catastrophic events occur, we have to allow that similar events may occur in the solar core, too, where all the ingredients are given. Namely, nonlinear plasma with its collective modes, magnetic field, and even a highly nonlinear energy source are present.

Based on a dynamical model of two-fluid equations with energy generation and loss, recently Li and Zhang [39] pointed out that, due to nonequilibrium of the solar core, coherent structures may develop and survive the destroying effects of frequent collisions and avoid decoherence at the high temperatures. Nuclear reactions go through inelastic collisions between ions and the accelerated ions transfer energy and momentum only subsequently via elastic collisions. The electrons lose energy and momentum through emitting photons via inelastic collisions with ions. On average, the direction of the energy and momentum flux in the nuclear fusion region is therefore: ions $\rightarrow$ electrons $\rightarrow$ photons. This implies that electrons do not reach thermodynamic equilibrium with ions ($T_i > T_e$), nor do photons with electrons in the fusion plasma. The non-equilibrium effect also shows itself in the two-fluid dynamical equations, and Li and Zhang have shown that the collision terms have been cancelled out by the non-equilibrium effect within general conditions. The reduced nonlinear two-fluid equations have been solved and it was shown that the self-organization of the stochastic thermal radiation field is in a close relation with the self-generated magnetic field: if one of them is absent, the other is very weak. Solving the model equations with a standard second-order explicit quasispectral method, they obtained that a magnetic field of at least $4 \cdot 10^4$ T may be generated at the centre of the Sun. Such a strong field shows that the self-organizing behaviour of the stochastic radiation field does occur. The growth time scale of the self-generated magnetic field is about $10^{12}$ s ($\approx 3 \cdot 10^4$ year).

Chang et al. [11] demonstrated that the sporadic and localized interactions of magnetic coherent structures arising from plasma resonances are the origin of 'complexity' in space plasmas. Dynamics of a plasma medium under the influence of a background magnetic field are characterized by coherent structures like convective cells, propagating nonlinear solitary waves, pseudo-equilibrium configurations and other varieties. They pointed out that MHD and kinetic (linear and nonlinear) instabilities result in fine-scale turbulences and trigger localized chaotic growth of a set of relevant order parameters. For typical MHD turbulence, the arising coherent structures are generally flux tubes. For our present purposes, it is important to notice that plasma interactions in general lead to sporadic and localized energy concentrations, and we suggest that the anisotropic dissipation of e.g. rotational energy of the solar core may enhance the energies of these sporadic and localized coherent structures significantly.

The localization of nuclear energy liberation needs a heating timescale $\tau_{nucl} = C_p T/(\varepsilon \nu)$ to be shorter than the timescale of the cooling processes $\tau_{expansion} = 5 \, H_p/v$, $\tau_{adj} = \kappa \rho^2 \, C_p R^2/(16 \sigma T^3)$ [30], where $C_p$ is the specific heat at constant pressure, $T$ the temperature of the heated region, $\varepsilon$ is the rate of energy liberation by nuclear reactions, $\nu$ is the exponent in the $\varepsilon \sim T^\nu$ relation, $H_p$ is the pressure scale height, 'v' the velocity of the heated region, $\kappa$ is a mean absorption coefficient, $\rho$ and $R$ is the density and the radius of the heated region, $\sigma$ is the Stefann-Boltzmann constant ($\sigma = 5,67 \cdot 10^{-8}$ W·m$^{-2}$·K$^{-4}$). With typical values ($\kappa = 0,2$ m$^2$·kg$^{-1}$, $\rho \sim 10^5$ kg·m$^{-3}$, $C_p = 30$ J·K$^{-1}$·mol$^{-1}$, $T = 10^7$ K,

$R = 10^4$ m, $H_p \sim 7 \cdot 10^7$ m, v $\sim 10$ - $10^4$ m/s), $\tau_{adj} \sim 7 \cdot 10^6$ s, $\tau_{expansion} \sim 3 \cdot 10^4$ - $10^7$ s, $\tau_{nucl} \sim 10^{16}$ s, while for $T = 10^8$ K, $\tau_{adj} = 7 \cdot 10^3$ s, $\tau_{expansion} \sim 3 \cdot 10^4$ - $10^7$ s, $\tau_{nucl} \sim 1$ s. It is clear that when the timescales of the cooling processes are comparable to or longer than the timescale of the rising motion of heated bubbles, the bubbles may travel significant distances. When the sporadic and localized energy enhancements heat a small macroscopic region above $10^8$ K, nuclear energy liberation will make the region explosive.

For the localization of the gravitational waves, Burgess et al. [7, 8] indicated the presence of density fluctuations in the deep solar core as a result of a resonant process similar to coronal heating. Energy is transferred from the g-modes into magnetic Alfven modes with density fluctuations corresponding to $\Delta T/T \sim 1,1$, dissipated heating energy is $Q_0 \sim 10^{28}$ J. Now our calculations [30] had shown that a much lower heating of $Q_0 \sim 10^{20}$ J is enough to produce a bubble rising a distance larger than its linear size.

Rotational, magnetic and thermal instabilities, gravitational g-mode oscillations, tidal effects and metastabilities make the solar core a dynamic, active system [27, 28, 30, 63]. The simple "ball of gas" concept, and the standard solar model based on this concept starts to lose its exclusive dominance in our picture of the Sun, and the emphasis may shift towards the dynamic phenomena of the Sun. A discrepancy between the old concept of the Sun and the accumulated results of the last decades makes it timely to present a revision of our concept of the Sun, and work out a deeper, more proper picture that may be useful for the paradigm shift from the "ball of gas" picture into the dynamic Sun framework.

In this paper, we present an approach, based on the concept that Sun has an astronomically large number of degrees of freedom (d.o.f.). There are mathematical methods worked out already to describe these dynamic collective modes offered by quantum field theories (e.g. [17, 18]); synergetics (e.g. [35]); self-organized criticality [3]; and so the approach suggested in this paper also opens up perspectives to new branches of solar sciences. We will shed some light on fields where the usefulness of this approach is indicated, and we will also propose some experiments to obtain measurable data on the nature of the Sun.

Sun is a self-organizing complex system. There are also some similarities to the most fundamental characteristics of terrestrial living organism. These lifelike characteristics correspond to the fact that the Sun shows a pronounced *activity, sensitivity, homeostasis, self-governance, integrality, and even a special type of "metabolism"* (obtaining energy from transforming materials into different compounds). Our aim is not to argue that the Sun is a living system or organism. Instead, our aim is to initiate a research to explore the most relevant facts, methods, and approaches suitable for the consideration of the nature of the Sun. We would like to make some initial steps in order to facilitate the exploration of the most important data, concepts and treatments necessary to work on this new field. To decide what is important for this aim we have to know what is a self-organising system, and what is life. In this introduction, we present a short preliminary consideration on the harder, and more fundamental question: on the lifelike nature of the Sun.

Three spectacular, easy-to-recognize signs of life are *activity, sensitivity and government from the level of the organism as a whole*. An exact study would necessitate a definition of the terms "activity", "sensitivity", and "government from the level of the organism as a whole". Instead, since our aim at present is only a stimulation of interest, we will postpone this more exact study to a later, more philosophical paper.

(i)   Is the Sun active? Yes, and its activity is known under the name: solar activity. The persistence of this activity adds a special emphasis on the significance of this activity, since a persistent activity is self-regenerating.

(ii)  Is the Sun a sensitive object? Yes. Every star is fundamentally sensitive, as nuclear energy production is an extremely sensitive function of temperature. Energy producing mechanisms are dependent on the high power of temperature ($\varepsilon_{pp} \sim T^4$ for the proton-proton cycle, and $\varepsilon_{CNO} \sim T^{20}$ for the CNO-cycle), a small amount of heating may lead to an acceleration of nuclear energy production and to further, increased amount of heating, which in turn may accelerate the heating process in a positive feedback cycle – if the negative feedback (heat loss processes) is weaker. This fundamental stellar sensitivity may be responsible for the development of local heated regions in stellar interiors and generating stellar activity from the radiative interior [30, 32].

(iii) Is solar activity regulated from the global level? Yes. "The prime cause of the solar cycle is a quasi-periodic oscillation of the SOLAR MAGNETIC FIELD." [42]. Solar cycle involves the sunspots, solar irradiance, surface flows, coronal shape, oscillation frequencies, etc. Sun is governed by a subtle agent (the magnetic field) that has a relatively small energy.

The Sun seems to govern its own behaviour from a high level of organization, and this would be a realization of a downward causation. The macroscopic degrees of freedom related to solar activity have significant energies even in comparison to the microscopic ones.

We note that the "bottom-up" approach building up the Sun from mass points or elementary particles is only one of the possible, physically sound approaches. For example, Eakins and Jaroszkiewicz [21, 22, 38] pointed out that a "top-down" approach offers a better description of our universe as a quantum universe, containing the bottom-up approach as a special, simplistic case. The quantum entanglement of the Sun to its global level and to the universe as a whole makes the Sun an even more highly complex system.

In this short paper we consider the nature of the Sun in an approach based on a consideration of the solar d.o.f., namely how the macroscopic d.o.f.s become excited, how the collective modes develop, and how they become organized by the magnetic field. The macroscopic degrees of freedom related to solar activity have significant energies even in comparison to the microscopic ones. The relative energy budget of solar activity to the average luminosity reaches an amplitude of 0,1 % [67, 68]. A whole series of phenomena related to intermediate collective degrees of freedom might play a role in the dynamics of the Sun that has yet escaped attention in the MHD approach.

## THE COLLECTIVE MODES OF THE SUN

Sound waves, MHD waves, p-mode, f-mode, g-mode, r-mode etc. oscillations, tidal waves, rising flux tubes etc. represent a continuous and self-regenerating activity extending throughout the whole body of the Sun and, in time, to a period comparable to the whole lifetime of the Sun. At a higher level of the solar collective modes, there appear the active regions, differential rotation, meridional circulation, giant, supergranular and granular activity, faculae, flares etc. At an even higher level we find the magnetic cycle with global collective modes and quasi-regularly repeating complex patterns of activity that represent such a long-lifetime and permanent self-initiating activity that they, apparently, do not have their counterpart in the realm of terrestrial physical self-organizing systems like sandpiles [3], lasers, Bénard-cells [35], phase transitions like ferromagnetism or superconductivity etc.

Appearance of Bénard cells is due to sharp temperature gradient. Different boundary conditions (rigid or free boundaries [10]) lead to different structural patterns of Bénard cells. The appearance of order is related to the macroscopic boundary conditions, which – being macroscopic - themselves represent macroscopic order that they can distribute to the system which they delimit. In that context, the phenomenon of Bénard cells is a member in the set of collective modes triggered by macroscopic physical influences. The phenomenon of generation of macroscopic order by the boundary conditions is a strange phenomenon of organizing remote control, organizing an astronomical number of microparticles into a coordinated behaviour to develop a coherent macroscopic order, in accordance with the boundary conditions. Del Giudice [19] pointed out that it is the electromagnetic radiative field, a long-range messenger indeed, that is responsible for such macroscopic organization. The quantum field formalism is also capable of describing the dynamism of the collective degrees of freedom [65].

In solid state physics collective phenomena found their description through the appearance of phonons and other quasiparticles. Plasma is very rich in collective phenomena, and our Sun consists largely from plasma. Actually, plasmons are suggested to describe collective plasma processes [44]. In addition to the haphazard individual fluctuations of electrons, Bohm saw there was a collective motion involving the electron sea as a whole, 'an electron plasma'. In such collective modes, the movement of individual electrons might appear to be 'random', but the cumulative effect of minute fluctuations in an enormous number ($10^{23}$) of electrons combined to produce an overall effect. Such collective effects representing a deeper order [6] were eventually well established experimentally and called 'plasmons' or, more generally, quasiparticles or excitons.

We know that the development of macroscopic modes received a rigorous foundation in the quantum theory of collective phenomena [53]. The macroscopic wave function gives a global collective description of the dynamical evolution in terms of a classical (but complex) wave field with a well-defined quantum phase. This is exactly what Saniga [49, 50] applied to describe the order represented in solar activity. He introduced a complex scalar Higgs field interacting with the electromagnetic gauge field in a modified Ginzburg-Landau theory. We suggest that the physical basis of the solar Higgs-fields is the development of collective modes of microscopic polarization effects, of electromagnetic quasiparticles contributing to "macromolecular dipoles" in the form of macroscopically polarized filaments. The solar Higgs field is a manifestation of quantum field changes related to dipole ordering, manifested also at the macroscopic degrees of freedom.

## IS THE SUN A SELF-ORGANIZING COMPLEX SYSTEM?

"Self-organized, non-equilibrium system … is a distinguishable collection of matter, with recognizable boundaries, which has a flow of energy, and possibly matter, passing through it, while maintaining, for time scales long compared to the dynamical time scales of its internal processes, a stable configuration far from thermodynamic equilibrium. This configuration is maintained by the action of cycles involving the transport of matter and energy within the system and between the system and its exterior. Further, the system is stabilized against small perturbations by the existence of feedback loops which regulate the rates of flow of the cycles." [55, p. 155].

Examples of natural self-organizing complex systems/phenomena are phase transitions like glaciation, magnetism, crystallization, pattern formation of snowflakes [25], avalanches of sandpiles [3], Benard-cells of convective flows, the generation of laser light [35], earthquakes in the mantle of the Earth. Applying the formulation of Bak [3, p. 5], we find that the Sun is a complex system, since it shows a definite and rich variability. By the concept of Haken [35, p. 11] the Sun is a self-organizing system, since a significant part of its collective modes are developing without a specific, impressing interference from the outside. Certainly, the Sun is able to change its internal structure through the homeostatic process of self-stabilization through its expansion and contraction, or otherwise. The Sun is a non-equilibrium system, which has a significant flow of energy passing through it while maintaining a stable configuration, having also a cyclic activity transporting energy and matter within the system and between the system and its exterior. The Sun is stabilized against small perturbations by the existence of negative feedback loops, since e.g. a heating perturbation generates expansion, and expansion causes cooling and return to the equilibrium, therefore fulfilling the criterion of Smolin [55, p. 155], too.

The solar surface is much more ordered than our present physical knowledge recognizes. Schwarzschild noticed in 1959 that granular structure should be turbulent instead of the observed cellular pattern. Reynolds number (the ratio of the inertial force of the motion to the decelerating drag force) characterizing the degree of turbulence is enormous – $Re \sim 10^{14}$. According to laboratory experiments, above $Re \sim 10^6$ the flow becomes chaotically turbulent, losing all its remaining regularities. Now the range $Re > 10^7$ is not yet reached experimentally. The extremely high Reynolds-number $Re \sim 10^{14}$ seems to indicate the necessity of a completely and extraordinarily chaotic turbulence of the solar granules. On the other hand observations show a quite different picture of an almost complete regularity, reminiscent to the cells of the beehives.

Solar activity is an expression of energy transfers from one mode to another, and to understand solar activity it is fundamental to keep in mind what are the relevant energy sources and actual modes of energy transfers.

**Table 1.** Estimated energies, luminosities and activity-related luminosity change amplitudes.

| Type | Energy, J | Luminosity, $J \cdot s^{-1}$ | Luminosity change amplitude, ($J \cdot s^{-1}$ in an $\sim$ 11 year cycle) |
|---|---|---|---|
| Nuclear | $2 \cdot 10^{44}$ | $4 \cdot 10^{26}$ | $4 \cdot 10^{23}$ |
| Gravitational | $4 \cdot 10^{41}$ | $3 \cdot 10^{22}$ | ? |
| Rotational | $2 \cdot 10^{36}$ | $6 \cdot 10^{19}$ | ? |
| Convective | $10^{31}$ | ? | ? |
| Magnetic | $5 \cdot 10^{27}$ | $2 \cdot 10^{19}$ | ? |
| Tidal | small | $< 10^{15}$ | $< 10^{15}$ |

The Earth [24, 40], galaxies [55] and the Universe [37, 54, 69] are already recognized to be complex self-organizing systems. Moreover, a comprehensive picture is proposed on cosmic evolution as the rise of complexity in nature [9]. There were attempts to apply self-organization in solar physics too. Li and Zhang [39] pointed out that coherent structures may be present in the solar core due to non-equilibrium processes.

Due to the self-organization of the stochastic radiation field, a self-generated magnetic field will develop in the solar core, the magnitude of which may reach $6 \cdot 10^4$ T.

Vlahos [66] regards the surface solar active regions as open dynamical systems away from equilibrium, driven by the turbulent convective zone. He has shown that all aspects of solar surface activity share one common characteristic, they are self-similar and their statistical behaviour follows well-defined power laws. Chang et al. [11] reviewed complexity, forced and/or self-organized criticality and topological phase-transitions in space plasmas.

## IS THE SUN A LIVING SYSTEM?

Living systems are such self-organizing systems that display high degree of functional self-governance of each organizational level [34]. The Sun is much more complex than ice, snowflake, sandpile, magnet, a Benard-cell, or a flame. *There is a significant difference between the most complex terrestrial complex systems (like Benard-cells or flames) and the Sun showing a lifelike activity involving thermodynamical, electromagnetic and gravitational behaviour on a wide range of spatial and temporal scales, degrees of freedom and energetic levels.* The special type of complexity of the Sun is related to dielectric macroscopic filaments and magnetic flux tubes, in a remarkable similarity to the terrestrial living organisms which are organized through the order of their polarized filamentary macromolecules also by electromagnetic fields [4]. The Sun is in a permanent state of activity and produces high-quality order in the patterns of solar activity [15], and in coherent sunlight [5, 45].

Ervin Bauer formulates three key marks of living systems [4, p. 32]. The first key mark says: "In order that a system could be alive, it should be able to show spontaneous changes even in an unchanging environment. Therefore it needs to have accumulated energy, which may be used within the conditions prevailing within the system and at unchanging outer conditions. This means that potential differences should set up within the system." The second key mark says: "The living systems show changes which are not due only to the effects of the outer changes but also to internal, non-mechanical factors…It is not necessary that the living systems react to every outer effect in every case with a work modifying the arising processes; our claim is only that they should not show only passive changes that can be determined unequivocally on the basis of the initial state of the system and the outer effect…The living systems show a sensitivity that is very wide-ranged, present to almost all outer effects, in the case of the repeated action of the outer effects as well. The reaction of the system is not a direct consequence of the outer effect, regarding its strength, too, and frequently the topographic agreement is also missing, when a stimulus acting on one place of the system corresponds to a reaction process occurring in a distant place of the system … the sensitivity of the living system is the ability to show processes that reacts to the equilibrating processes elicited by the outer effects by other processes using energy to recharge the potential differences" [4, pp. 36-42]. Therefore, the persistence of such sensitivity involves a self-regenerative activity, recharging the most important living potentials in the most suitable way. Such a self-regeneration assumes the presence of free energy available for such a work. Now if the sensitivity is present in most of the places of the system, and if it may act non-locally, this requirement apparently necessitates a free energy reservoir that is available by many or most parts of the system. Therefore, the free energy supply of almost any part of the system at any time assumes the presence of an activity that mobilizes and directs the free energy to the proper place in due time. This governing activity should consume much less energy than the system's "real activity" that concerns to the outer effects and the corresponding internal changes. In this way the global sensitivity of living systems seems to involve the presence of a subtle free energy field that is able to govern the processes developing from the stimulus until the answer in a functionally reasonable manner. Now the question is how is it that such a governing subtle energy field is present in the Sun, and how it is able to govern from "above", from the global level of the living system.

Such a governance is enlightened by Ervin Bauer with his third requirement: "The work of the living system, independently of the environmental conditions, is used against the equilibrium towards which the system would proceed by the physico-chemical laws at the actual outer conditions and the initial state of the system." Ervin Bauer formulated the life principle: "Living systems and only living systems are never in a state of equilibrium; to the debit of their reserves of free energy they continually perform work against the setting in the equilibrium state which should, according to the laws of physics and chemistry, be established under the actual external conditions" [4, p. 44].

The fact that life is intimately related to collective phenomena is an inevitable necessity. Microparticles have only individual degrees of freedom. Propensities [47] act as generalized determinations. They are situation-dependent collective phenomena developing in many-sided, dynamic, changing situations. Propensities are already related to collective degrees of freedom. Living systems are possible only above a certain level of complexity because the appearance of collective modes depends critically on the level of complexity. The higher the complexity, the richer and more field-like the propensities will be. Living systems may develop when propensities are so complex that they develop a field-like character that coheres with an energy reservoir available for this field. Of course, this statement is true only together with its dual: living systems may develop only when the life principle meets with a system complex enough that a field-like causative structure may develop with a field-like energy reservoir.

Regarding the electromagnetic nature of the solar cycle, it is important to be aware to the work of Del Giudice [19]. He recognized that a missing ingredient of liquid water formation as condensed matter from water vapour is the electromagnetic (e.m.) radiative field. In the process of water condensation, water density increases abruptly by a factor of 1600 at the condensation point of 373 K, and this process occurs remarkably fast regarding the fact that more than $10^{23}$ individual water molecules have to be hooked together almost instantaneously. Although the contribution of the e.m. radiative field to the interaction between two particles is exceedingly small with respect to the static forces, but, when we collect a very large number $N$ of components, it is possible to realize [20] that at a density exceeding a threshold the radiative contribution becomes very large (superradiance). In the solar interior, the plasma and its filamentary structures interacting with the strong radiative field may produce a coherent phase (in the context of radiative field) or collective mode (in the context of the plasma constituents), similarly to water. The coherent phase represents only a fraction of the solar plasma, and so a two-fluid picture emerges, again resembling the picture of the superfluid helium. At the same time, the coherent field is able to arrange an astronomical number of plasma particles almost instantaneously when suitable conditions develop. We know that the electrodynamic properties of water depend on the coherent phase, and, similarly, we can expect that the electrodynamic properties of the solar plasma also depend on the coherent collective mode. In the process of formation of coherent collective modes, a sizeable polarization field develops, which may contain a significant amount of information [19], and hooks together an astronomical number of particles by the coupling of quantum electrodynamic fields. Indeed, it is well-known that water may show a complex behaviour, it has memory, structure, and information (e.g. [14]). Similarly, the formation of a filamentary basic plasma structure in the solar core is indicated by Li and Zhang [39]; and in the solar convective zone [49, 52] and in space plasma by Chang et al. [11]. On this basis, we think that the detailed calculations of Del Giudice and his group may substantiate that the solar plasma may be rich enough in information to make the Sun capable of maintaining its lifelike organisation.

The fellow systems of the Sun, the Earth (e.g. [40, 64]), the Galaxy [33] and the Universe [22] were already considered as living systems/organisms. Recently, Grimm [34] concluded that Gaia System is an authentic living system. The idea that a galaxy is a self-organized system – more an ecology than a nonliving clump of stars and gas – has become common among astronomers and physicists who study galaxies [55]. Some regards galaxies as literally alive: "I have argued that our Galaxy is alive – literally alive, in the full biological meaning of the term. … The striking feature of the way in which spiral galaxies maintain a steady state, far from equilibrium … has been produced by a process of evolution and competition" [33, p. 214]. We note that these approaches are merely physical ones, and as such, they may miss the most substantial points of living organisms. "From the standpoint of the traditional physical sciences a living organism appears as a complex aggregation of separable units of matter associated causally with one another in separable events. From the biological standpoint, on the other hand, the apparently separable units of matter and events are seen not to be actually separable, but, in their relationships to one another, to be taking part in the manifestation of the co-ordinated and persistent whole which is called a life, and which has no spatial limits. Thus the separable units and events of physical interpretation are seen to be illusory, but the same phenomena become intelligible in so far as they can be interpreted as phenomena of life" [36, p. 62]. "Physical interpretation, in so far as we adhere to it, is applicable to the whole of our perceived experience. But so is biological interpretation … Inasmuch, however, as in biological interpretation we are taking our experience more fully into account than in physical interpretation, biological interpretation is on a higher level, and represents reality less incompletely than physical interpretation." [36, p. 64]. The biological standpoint is worked out in our books [29, 31].

## THERMODYNAMIC CHARACTERISTICS OF THE INEQUILIBRIUM SUN

The Sun has much more degrees of freedom and much higher energy flux than the surface of the Earth, therefore the Sun shows much higher versatility, plasticity and inner drive towards autonomous activity than systems at the surface of the Earth. An important quantitative characteristic of any complex and living system is the distance from thermodynamic equilibrium. The distance from the thermodynamic equilibrium can be measured with extropy.

Martinás [41, p. 39] calculated the extropy flow $J_\Pi = L_2/T_2 - L_1/T_1$ (where $L$ is the luminosity, $T$ is the temperature, index 1 refers to the incoming and index 2 to the outgoing values) for the Earth as being $J_\Pi(\text{Earth}) = 4 \cdot 10^{14}$ J K$^{-1}$ s$^{-1}$. We calculated the extropy flow for the Sun: $J_\Pi(\text{Sun}) \sim L(Sun)/T_{\text{eff}}(\text{Sun}) = 7 \cdot 10^{22}$ J K$^{-1}$ s$^{-1}$. This number indicates that the Sun is a strongly nonequilibrium system. Regarding that the numerical values of extropy flows are generally not really familiar, we calculated also some representative values for comparison. For an (unmoving) human the heat output is $L \sim (2$ J·kg$^{-1}$ s$^{-1}$ $7 \ 10^5) \cdot g = 1{,}4 \cdot 10^3$ J s$^{-1}$ as estimated on the rate of metabolism. Therefore, for an unmoving human: $J_\Pi(\text{human}) \sim L/T = 4$ J K$^{-1}$ s$^{-1}$. The extropy flow for unit masses ($j_\pi$) is $j_\pi$(Sun) = 3,5·$10^{-14}$ J $K^{-1}$ $s^{-1}$·$kg^{-1}$, $J_\pi$(Earth) = 6,7·$10^{-17}$ J $K^{-1}$ $s^{-1}$·$kg^{-1}$, and $J_\pi$ (human) = 5,7·$10^{-2}$ J $K^{-1}$ $s^{-1}$·$kg^{-1}$. In the solar emitting layer, the photosphere, $J_\pi$(photosphere) ~ 14 · J $K^{-1}$ $s^{-1}$·$kg^{-1}$. We do not think that one can immediately draw definite conclusions on the basis of these numbers. Nevertheless, if other facts will point to the same direction, these results may indicate that solar photosphere is more favourable for lifelike complexity than the surface of the Earth.

**Table 2.** Some estimated specific extropy flows.

| Generic structure | Average $j_\pi$, J $K^{-1}$ $s^{-1}$·$kg^{-1}$ |
|---|---|
| Stars | $10^{-8} - 10^{-4}$ |
| Earth | 7·$10^{-11}$ |
| Prebiotic material | 3·$10^{-5}$ |
| Human body | 6·$10^{-2}$ |
| Solar photosphere | 10 |

Chaisson [9, p. 139] presented the free energy rate for the modern society a $\Phi_m$ value 50 J·$s^{-1}$·$kg^{-1}$, that is $j_\pi$ = 2 J $K^{-1}$ $s^{-1}$·$kg^{-1}$, this number arises if we count all the energies used by man but the related mass is the mass of mankind. When all the mass of these equipments (and the mass of the equipments producing these ones etc.) is involved, then the value decreases by some orders of magnitude. In this way, modern society loses its 'top position'.

## EXPERIMENTS SUGGESTED TO TEST THE NATURE OF THE SUN

The Sun has a life favouring effect through the negentropy (extropy) it carries, transporting free energy available for living organisms to do work. The Sun plays a life-inducing role in the earth [45, p. 143]. Popp and Yan [46] demonstrated (see their Figs. 8a and 8b) that spontaneous emission of biological systems may also originate from squeezed states as it is shown by their emissions, a significant part of which had fallen below the Poissonian statistics $p(0) < e^{-<n>}$. In case of artificial light sources, the emitted radiation never had fallen below the Poissonian photocount statistics. We suggest that this method may be applied to test the nature of solar electromagnetic radiation. If the solar radiation contains non-classical squeezed light, the photocount statistics may be similar to the spontaneous emissions of biological systems, and this would be an important element to understand the emergence of life. The apparent fact is that greenhouse vegetables have less natural aroma and nutritive, vitalizing power than the same vegetables grown on open air and sunshine.

Our study proposes to consider if there are any life-enhancing solar effects besides the extropy carried in solar radiation. We propose that one candidate from the wide range of solar effects is the information content of solar radiation. Recently, Consolini et al. [15] have shown by means of the normalized information entropy measure that the regularities of granular pattern present evidence of a spatial-temporal organization in the evolution of convective pattern. We point out that this and/or other types of information present in solar phenomena (e.g. in the Higgs-field [51, 52]) may also carry life-enhancing effects. We suggest the experimental study of the growth of different organisms (bacteria, plants, eggs) on the influence of different types of light: (i) room-temperature artificial light sources; (ii) high-temperature (visible, UV, EUV) artificial light; artificial light with the same temperature as that of the sunlight $T$ = 5 778 K (S1); artificial light with the same spectra as sunlight (S2); artificial light with the same spectra, time and incidence angle variation as that of sunlight (S3)

(iii) artificial electromagnetic waves like the waves of radio-broadcasts and television-broadcast; informatically modulated artificial lights (like S1, S2, S3); (iv) and light from living organisms, biophotons. In comparison, one may study (v) the biological influence of natural solar light. These experimental possibilities would give a better understanding of the nature of Sun.

## CONCLUDING REMARKS

The present approach aims to build up a systematic picture on the network of collective and coherent modes of the Sun, and offers a higher resolution of the physical and possible lifelike nature of the Sun. The obtained results draw some methods and formalisms into solar physics like complexity science, quantum field theory, information theory, theoretical biology, biophoton research, synergetics, condensed matter physics, self-organization and non-linear sciences etc., some of which actually play a central role in the dynamism of modern sciences.

## ACKNOWLEDGMENTS

The author expresses his thanks to Dr. Metod Saniga for the encouragement, for the consultations and for the suggestions made to improve the original manuscript; and to Jean Drew for improvements of the English. We also express our thanks to Dr. Katalin Martinás for her remarks and suggestions.

## REFERENCES

[1]   Allen, C.W.: *Astrophysical Quantities*. 3rd edition.
       University of London, The Athlone Press, 161, 1963,

[2]   Bahcall, J.N.: *Neutrino Astrophysics*.
       Cambridge University Press, 43, 1989,

[3]   Bak, P.: *How Nature Works. The Science of Self-Organized Criticality*.
       Copernicus, New York, 1996,

[4]   Bauer, E.: *Theoretical Biology*.
       VIEM, Moscow, 1935; Akadémiai Kiadó, Budapest, 1967,

[5]   Bischof, M.: *Biophotonen-das Lichts in unseren Zellen*. 11th edition.
       Zweit+sendeins, Frankfurt, 2001,

[6]   Briggs, J.P. and Peat, F.D.: *Looking Glass Universe*.
       Fontana Paperbacks, Glasgow, 101-102, 1984,

[7]   Burgess, C. et al.: *Large Mixing Angle Oscillations as a Probe of the Deep Solar Interior*.
       Astrophysical Journal **588**, L65-L68, 2003,

[8]   Burgess, C.P. et al.: *Resonant origin for density fluctuations deep within the Sun: helioseismology and magneto-gravity waves*.
       IFIC-03-12; McGill-02/40
       http://arxiv.org/abs/astro-ph/0304462,

[9]   Chaisson, E.: *Cosmic Evolution. The Rise of Complexity in Nature*.
       Harvard University Press, Cambridge, 139, 2001,

[10]  Chandrasekhar, S.: *Hydrodynamic and Hydromagnetic Stability*.
       Clarendon Press, Oxford, 1961,

[11]  Chang, T.; Tam, S.W.Y.; Wu, C.-C. and Consolini, G.: *Complexity, Forced and/or Self-Organized Criticality, and Topological Phase Transitions in Space Plasmas*.
       Space Science Reviews **107**, 425-445, 2003,

[12] Charbonneau, P. and MacGregor, B.: *Solar spin-down with internal magnetic fields*.
Astrophysical Journal **397**, L63-L66, 1992,

[13] Charbonneau, P. and MacGregor, B.: *Solar Spin-down with Internal Magnetic Fields: Erratum*.
Astrophysical Journal **403**, L87, 1993,

[14] Chaplin, M.: *Water Structure and Behavior*.
http://www.lsbu.ac.uk/water,

[15] Consolini, G. et al.: *Information entropy in solar atmospheric fields I. Intensity photospheric structures*.
Astronomy and Astrophysics **402**, 1115-1127, 2003,

[16] Couvidat, S. et al.: *The Rotation of the Deep Solar Layers*.
Astrophysical Journal **597**, L77-L79, 2003,
http://xxx.lanl.gov/abs/astro-ph/0309806,

[17] Del Giudice, E.; Doglia, S.; Milani, M. and Vitiello, G.: *A quantum field theoretical approach to the collective behaviour of biological systems*.
Nuclear Physics B**251**, 375-400, 1985,

[18] Del Giudice, E.; Doglia, S.; Milani, M. and Vitiello, G.: *Electromagnetic field and spontaneous symmetry breaking in biological matter*.
Nuclear Physics B**275**, 185-199, 1986,

[19] Del Giudice, E.: *Collective processes in living matter: A key for homeopathy*.
Homeopathy in Focus. Verlag für Ganzheits Medizin, Essen, 14-17, 1990,

[20] Dicke, R.H.: *Coherence in spontaneous radiation processes*.
Physical Review **93**, 99-110, 1954,

[21] Eakins, J. and Jaroszkiewicz, G.: *The Quantum Universe*.
http://arxiv.org/abs/quant-ph/0203020,

[22] Eakins, J. and Jaroszkiewicz, G.: *The origin of causal set structure in the quantum universe*.
http://arxiv.org/abs/gr-qc/0301117,

[23] Elgin, D.: *Our living universe*.
http://www.noetic.org/publications/review/issue54/r54_Elgin.html,

[24] Ghil, M.: *The Earth as a complex system*.
3rd IGBP Congress, Connectivities in the Earth System, Banff, Canada, 19-24 June 2003,
http://www.igbp.kva.se/congress/abstracts/Ghil_congress_abstract.pdf,

[25] Gleick, J.: *Chaos. Making a New Science*.
Cardinal, Bungay, Suffolk, 310, 1988,

[26] Goossens, M.: *An Introduction to Plasma Astrophysics and Magnetohydrodynamics*.
Kluwer Academic Publishers, Dordrecht/Boston/London, 1, 2003,

[27] Grandpierre, A.: *How is the sun working*?
Solar Physics **128**, 3-6, 1990,

[28] Grandpierre, A.: *A pulsating-ejecting solar core model and the solar neutrino problem*.
Astronomy & Astrophysics **308**, 199-212, 1996,

[29] Grandpierre, A.: *The Book of the Living Universe*. In Hungarian.
Válasz Könyvkiadó, Budapest, 2002,

[30] Grandpierre, A., *On the Development of Thermal Metastabilities in the Solar Core*.
Astrophysics and Space Science, submitted,

[31] Grandpierre, A.: *The Book of the Living Universe*.
in preparation,

[32] Grandpierre, A. and Agoston, G.: *On the onset of thermal metastabilities in the solar core.*
Astrophysics and Space Science, accepted,

[33] Gribbin, J.: *In the Beginning. The Birth of the Living Universe.*
Penguin Books, London, 1993,

[34] Grimm, K.A.: *Is Earth a Living System*?
2003 Seattle Annual Meeting of The Geological Society of America, (2.-5. XI. 2003),
http://gsa.confex.com/gsa/2003AM/finalprogram/abstract_62123.htm,

[35] Haken, H.: *Information and Self-Organization*. 2nd Enlarged Edition.
Springer, Berlin, 2000,

[36] Haldane, J.S.: *The Philosophy of a Biologist.*
Clarendon Press, Oxford, 1935,

[37] Jantzsch, E.: *The Self-Organizing Universe.*
Pergamon Press, Oxford, 1980,

[38] Jaroszkiewicz, G.: *The running of the Universe and the quantum structure of time.*
http://arxiv.org/abs/quant-ph/0105013,

[39] Li, L.H. and Zhang, H.Q.: *The dissipative effect of thermal radiation loss in high-temperature dense plasmas.*
Journal of Physics D: Applied Physics **29**, 2217-2220, 1996,
http://arxiv.org/abs/astro-ph/9711002,

[40] Lovelock, J.: *The Ages of Gaia: A Biography of our Living Earth*, OUP, Oxford, 1988. New edition, 1996,

[41] Martinás, K.: *The Entropic Limits of the Possible Futures.*
Economic and State University, Budapest, 2002,

[42] Ossendrijver, M. and Hoyng, P.: *Solar Cycle.*
Murdin, P., ed.: *Encyclopedia of Astronomy and Astrophysics*. Institute of Physics Publishing, Bristol and Philadelphia; Nature Publishing Group, London, New York and Tokyo, 2502, 2001,

[43] Peratt, A.L. *Physics of the Plasma Universe.*
Springer-Verlag, New York-Berlin, 19, 1992.

[44] Pines, D.: *Collective Energy Losses in Solids.*
Reviews of Modern Physics **28**, 184-198, 1956,

[45] Popp, F.-A.: *Biologie des Lichts. Grundlagen der ultraschwachen Zellstrahlung.*
Verlag Paul Parey, Berlin and Hamburg, 1984,

[46] Popp, F.A. and Yan Y.: *Delayed luminescence of biological systems in terms of coherent states.*
Physics Letters A **293**, 93-97, 2002,
http://www.lifescientists.de/publication/pub2001-07.htm,

[47] Popper, K.: *A World of Propensities.*
Thoemmes, Bristol, 1990,

[48] Ridpath, I., ed.: *A Dictionary of Astronomy.*
Oxford University Press, Oxford, New York, 1997,

[49] Saniga, M.: *On the remarkable similarity between the sunspot and the type II superconductor magnetic vortex*. Ph.D. Thesis.
Slovak academy of Sciences, Bratislava, 1990,

[50] Saniga, M.: *A Sunspot as the Macroscopic Analog of a Magnetic Vortex in a Type II Superconductor.*
Soviet Astronomy **36**, 466-468, 1992,

[51] Saniga, M. and Klačka, J.: *Quantum Micro-Solitons – A Clue to Solve the Fundamental Problems of Solar Physics*?
Astrophysics and Space Science **200**, 1-7, 1993,

[52] Saniga, M.: *On the Possibility of a Chern-Simons Physics on the Sun*.
Chaos, Solitons & Fractals **7**, 1053-1055, 1996,
http://www.ta3.sk/~msaniga/pub/solar-physics.html,

[53] Sewell, G.L.: *Quantum Theory of Collective Phenomena*.
Clarendon Press, Oxford, 1986,

[54] Shermer, M.: *Digits and Fidgets*.
Scientific American, January 2003,

[55] Smolin, L.: *The Life of the Cosmos*.
Oxford University Press, New York, Oxford, 1997,

[56] Spiegel, E.A. and Zahn, J.-P.: *The solar tachocline*.
Astronomy & Astrophysics **265**, 106-114, 1992,

[57] Spruit, H.C. and Knobloch, E.: *Baroclinic instability in stars*.
Astrophysical Journal **132**, 89-96, 1984,

[58] Spruit, H.C.: *Dynamo action by differential rotation in a stably stratified stellar interior*.
Astronomy & Astrophysics **381**, 923-932, 2002,

[59] Talon, S.; Kumar, P. and Zahn, J.-P.: *Angular Momentum Extraction by Gravity Waves in the Sun*.
Astrophysical Journal **574**, L175-L178, 2002,

[60] Tayler, R.J.: *The adiabatic stability of stars containing magnetic fields-. Toroidal fields*.
Monthly Notices of the Royal Astronomical Society **161**, 365-380, 1973,

[61] Tsytovich, V.N.: *Nonlinear Effects in Plasma*.
Plenum Press, New York and London, 1, 1970,

[62] Turck-Chieze, S.; Nghiem, P.; Couvidat, S. and Turcotte, S.: *Solar Internal Composition and Nuclear Reaction Rates in the Light of Helioseismology*.
Solar Physics **200**, 323-342, 2001,

[63] Turck-Chieze, S.:
Murdin, P., ed.: *Encyclopedia of Astronomy and Astrophysics*. Vol. 3, IOP Publ. Bristol, 2616, 2001,

[64] Vernadsky, V.I.: *The Biosphere*.
1926; Copernicus - Springer-Verlag, New York, 1997,

[65] Vitiello, G.; Del Giudice, E.; Doglia, S. and Milani, M.:
Fröhlich, H. and Kremer, F., eds.: *Coherent Excitations in Biological Systems*. Springer, New York, 469, 1983,

[66] Vlahos, L.: *Statistical properties of the Evolution of solar magnetic fields*.
Sawaya-Lacosta, H., ed.: SOLMAG 2002, ESA Publ. Div., 105, 2002,

[67] Walton, S.R.; Preminger, D.G. and Chapman, G.A.: *The Contribution of Faculae and Network to Long-Term Changes in the Total Solar Irradiance*.
Astrophysical Journal **590**, 1088-1094, 2003,

[68] Willson, R.C. and Hudson, H.S.: *Solar luminosity variations in solar cycle 21*.
Nature **332**, 810-812, 1988,

[69] Zizzi, P.A.: *Quantum Computation toward Quantum Gravity*.
General Relativity and Gravitation **33**, 1305-1318, 2001,
http://arxiv.org/abs/gr-qc/0008049.

# KONCEPTUALNI KORACI PREMA ISTRAŽIVANJU PRIRODE NAŠEG SUNCA

Attila Grandpierre

Zvjezdarnica Konkoli
Budimpešta, Madžarska

## SAŽETAK

Jedno od temeljnih pitanja u solarnim istraživanjima je o prirodi Sunca. U ovom radu pokazujemo da stanje plazme Sunca dovodi do samorazvijanja solarne aktivnosti. Oslobađanje magnetske, rotacijske, gravitacijske, nuklearne energije i energije gravitacijskih vibracija uzrokuje odstupanja od uniformnosti i sferne simetrije. Putem nestabilnosti to dovodi do nastajanja sporadičnih i lokaliziranih područja poput cijevi toka, električkih ispuna, magnetskih elemenata i područja visokih temperatura. Prikazan je sustavni pristup istraživanju solarnih kolektivnih stupnjeva slobode, do uključivo pojave reda zbog magnetskih svojstava povezanih s Higgsovim poljima.

Razmatranje solarne aktivnosti kao pretvorbe oblika energija vodi na sliku mreže energetskih razina Sunca, pokazujući kako Sunce nije ni obična kugla plina, niti stacionarni fuzijski reaktor, nego složeni, samoorganizirajući sustav. Budući da su složeni, samoorganizirajući sustavi slični živim sustavima (a po nekim shvaćanjima identični njima) također razmatramo argumente koji ukazuju na živu prirodu Sunca. Termodinamička svojstva neuravnoteženog Sunca su u tom smislu značajna. Numerički su procjenjene gustoće stope slobodne energije i specifične eksergije.

## KLJUČNE RIJEČI

solarna fizika, stupnjevi slobode, samoorganizirajući složeni sustavi, neravnotežna termodinamika, astrobiologija